\documentstyle[12pt]{article}
\author{Klaus Lucke\thanks{Present adress: DAMTP ,University of Cambridge,
Cambridge, U.K.} \\ Institut f\"ur Theoretische Physik III\\ Universit\"at
Erlangen \\ Staudtstr. 7 \\D-91058 Erlangen}
\title{Regularization Dependence of the Zero Mode Dynamics in the Schwinger
Model}

\begin{document}
\maketitle
\abstract

I compare heatkernel regularization with sharp gauge invariant cutoffs in the
Hamiltonian formulation of the Coulomb gauged Schwinger model on a circle. The
effective potential for the zero mode of the gauge field in a given fermionic
configuration is different in these two regularizations, the difference being
independent of the chosen fermionic configuration. In the continuum limit $L
\rightarrow \infty$ the gauge field can be localized or delocalized depending
on the regulator.

\section{Introduction}

In QFT singularities in a continuum formulation force us to introduce a
regularization procedure to extract finite physical quantities and to give a
clear definition of the theory \cite{itzykson}.

We are immediately faced with the question whether the resulting predictions
will depend on the regularization procedure. In gauge theories it is necessary
to demand that the regulator respects the gauge symmetry. In the Schwinger
model (and other models!) the anomaly can be missed when gauge invariance is
not manifestly kept. Generally it is conjectured that, as long as the symmetry
requirement is met, results are independent of the regulator, and hence the
most convenient procedure is chosen.

In this article I show that this assumption is not always justified. I discuss
different gauge invariant regularizations that lead to different predictions.

The article is organized as follows: I give a brief introduction to the
Schwinger model in section 2. In section 3 I compare different regularizations
keeping the perimeter $L$ of the circle finite. Section 4 is devoted to the
limit $L \rightarrow \infty$.

\section{Schwinger model}

The Schwinger model is QED in $1+1$ dimensions with vanishing quark mass. As it
can be completely solved \cite{manton}, it serves for testing new techniques
and questions of principle. The Lagrangian is given by
$${\cal L} = \bar \psi i /\!\!\!\!D \psi -\frac{1}{4}F^{\mu \nu} F_{\mu \nu}
\qquad .$$
On a circle with periodicity condition for $A_\mu$ and antiperiodicity for the
two component spinor $\psi$ we can identify our true degrees of freedom in the
Coulomb gauge $\partial_1A_1=0$ as the fermionic field $\psi$ and the
(dimensionless) zero mode of the gauge field
$$c=\frac{g}{2 \pi}\int_0^L A_1(x) dx +\frac{1}{2} \qquad .$$
$c$ cannot be gauged away because of the boundary condition for the fermions.
The only left over gauge symmetries are large gauge transformations:
$$\psi \rightarrow \psi e^{2 \pi i n \frac{x}{L}} \qquad , \qquad c \rightarrow
c+n \qquad .$$
For the calculation of the Hamiltonian we have to express the non-dynamical
$A_0$ in terms of the dynamical variables using the Lagrange equation for
$A_0$.
We impose the charge neutrality condition on phy\-si\-cal states to allow for a
solution of this equation. In terms of the dimensionless Fourier modes of the
fermionic field in the chiral representation
$$\left( \begin{array}{c}
	a_n \\ b_n
	\end{array}
\right) = \frac{1}{\sqrt{L}}\int_0^L dx e^{2 \pi i (n+\frac{1}{2}) \frac{x}{L}}
\psi (x)$$
the Hamiltonian reads
\begin{eqnarray}
H&=&-\frac{g^2 L}{8 \pi^2} \frac{\partial^2}{\partial c^2} + \frac{2 \pi}{L}
\sum_n (n+c)\Bigl(R(-(n+c)) b_n^\dagger b_n-R(n+c) a_n^\dagger a_n\Bigr)
\nonumber\\
&& +\sum_{n\not=0} \frac{g^2 L}{8 \pi^2 n^2} j_n^\dagger j_n \nonumber
\end{eqnarray}
with $j_n=\sum_m a_m^\dagger a_{m+n}+ b_m^\dagger b_{m+n}$ .
The first term represents the kinetic energy of the zero mode of the gauge
field, while the second term accounts for the non-interacting part of the
fermionic energy mi\-ni\-mal\-ly coupled only to the zero mode. The third term
is the Coulomb interaction resulting from all terms originally involving $A_0$
and is not important in the following discussion. I have introduced the
function $R$ for regularization. For the moment we ought to (naively) think
$R\equiv 1$, in order to identify the Dirac sea. Note that the Hamiltonian is
automatically gauge invariant (under residual gauge transformations) for any
function $R$, as $c$ and $n$ only appear in the combination $c+n$ and the sum
is over all integers.

Clearly the second term forces us to construct a Dirac sea $|0>$ with modes
$a_{n}$, $b_{-n}$ occupied (empty) for $n\geq0$ ($n<0$) as a re\-fe\-rence
state. The Hilbert space is spanned by (charge neutral) states obtained by
acting on $|0>$ with a finite number of $b$'s, $b^\dagger$'s, $a$'s and
$a^\dagger$'s. Neglecting the Coulomb interaction, these states are eigenstates
of the fermionic part of the Hamiltonian and hence have a well defined
``energy'', resulting in an effective Hamiltonian for the zero mode $c$. In the
following I discuss the regularization of this energy for the reference state
$|0>$. For other states the discussion can be generalized easily.

\section{Comparison at finite L}

The regulator $R$ is used as a weight for modes.  If $R=1$, a mode is kept. For
$R=0$ the contribution from this mode is discarded.
In order to obtain a reasonable theory $R$  has to satisfy
$$R(u)\approx 1 \qquad \forall \mbox{ moderate } u$$
$$R(u)\approx 0 \qquad \forall u \gg 1 \qquad .$$
In this section I define and discuss two special regulators, heatkernel
regularization and the sharp cutoff. The sharp cutoff includes a free parameter
(the function $f$) and is hence rather a family of cutoffs. We will see that
the results crucially depend on the parameter $f$. The regulators are defined
by ($\Lambda \rightarrow \infty$ at the end of the calculation)
$$ R^{heat}(u):=e^{-\frac{u}{\Lambda}}$$
\medskip
$$
R^f(u) := \left\lbrace \begin{array}{ccc}
		1 & \qquad &u<\Lambda-1\\
		f(u-\Lambda)&&\Lambda-1<u<\Lambda\\
		0&&\Lambda<u
	\end{array}  \right . \qquad .
$$
$f$ parametrizes how the discontinuity of the step is smeared out (over one
single mode), where we assume $f(0)=0$, $f(-1)=1$, $f([-1,0])=[0,1]$ and $f$
smooth.
Discarding constants, the energy of the Dirac sea (without Coulomb interaction)
is calculated in these regularizations to be (as an operator w.r.t. the wave
function $|\phi>$ for the zero mode):
\begin{eqnarray}
H^{heat}|0>\otimes |\phi>&=& |0>\otimes \Bigl(-\frac{g^2 L}{8
\pi^2}\frac{\partial^2}{\partial c^2}+\frac{2 \pi}{L}(c-\frac{1}{2})^2
\Bigr)|\phi> \nonumber\\
H^f|0>\otimes |\phi>&=& |0>\otimes\Biggl(-\frac{g^2 L}{8
\pi^2}\frac{\partial^2}{\partial c^2}\nonumber\\
&+&\frac{2 \pi}{L}\biggl(\Lambda \Bigl(1-f(-1+\tilde c)-f(-\tilde c)\Bigr)
+\mbox{finite}\biggr) \Biggr)|\phi>
\nonumber
\end{eqnarray}
$\tilde c \in ]0,1[ $ is defined to be the non-integer part of $c$
($c=[c]+\tilde c$) and hence gauge invariant \cite{foot}.

\medskip

{\bf Case 1: $1-f(-\tilde c)-f(-1+\tilde c) \not\equiv 0$}\\
The potential is singular for $\Lambda \rightarrow \infty$ and dominated by the
periodic (because it is not dependent on $[c]$) term $\Lambda (1-f(-\tilde
c)-f(-1+\tilde c))$. So the only finite-energy-eigenstates $|\phi>$ are
localized in the minima of the singular potential. Tunneling between the (gauge
equivalent) minima is impossible as the potential barrier is infinite.

\medskip

{\bf Case 2:  $1-f(-\tilde c)-f(-1+\tilde c) \equiv 0$}\\
This exceptional case is completely regular. Without loss of generality we can
select $f_0(x):=1+x$.
Other choices yield slightly different results, but no spectacularly new
feature can occur.
For comparison with the heatkernel regularization we also look at the finite
parts:
$$H^{f_0}|0>\otimes |\phi>= |0>\otimes\Biggl(-\frac{g^2 L}{8
\pi^2}\frac{\partial^2}{\partial c^2}+\frac{2 \pi}{L}
\Bigl((c-\frac{1}{2})^2+(\tilde c-\frac{1}{2})^2\Bigr)\Biggr)  |\phi>$$
This differs from the heatkernel regularization by the dependence on $\tilde
c$.

\medskip
Hence for finite $L$ we have a variety of effective Hamiltonians for the zero
mode; each describing different dynamics.
As this is rather unusual, a brief remark on gauge invariance is necessary. The
Hamiltonian was regularized in a manifestly gauge invariant manner. On the
other hand our reference state is not gauge invariant. However as the
calculation is basically independent of the reference state, we could equally
have taken a gauge invariant linear combination of states ($\theta$-vacua). We
would still end up with regulator dependence as the dependence on $\tilde c$ is
the same  for $\theta$-vacua. Hence we can continue discussing just $|0>$ .

\section{Continuum limit}

Performing the continuum limit in  the sense of a strong coupling limit $gL
\rightarrow \infty$, we obtain that the kinetic energy term will be dominant
for heatkernel regularization and case-2-sharp-cutoffs. Hence the low lying
(finite energy) states will be very much {\it delocalized}. Thus all dependence
on $\tilde c$ can be safely replaced by a constant and all results are
equivalent.

On the other hand we have to be careful about the order of limits for
case-1-sharp-cutoffs.  If we perform the {\it regulator limit before} (after)
the continuum limit, we will obtain a {\it localized} (delocalized) zero mode.
More precisely we obtain for case 1
$$ \mbox{delocalization for} \qquad \frac{g^2L^2}{\Lambda} \rightarrow \infty$$
$$ \mbox{localization for} \qquad \frac{g^2L^2}{\Lambda} \rightarrow 0 \qquad
.$$
Note that $\Lambda$ is a dimensionless cutoff. In terms of a physical
momen\-tum cutoff $p_{cut}=\frac{2 \pi \Lambda}{L}$ the crucial ratio is
(disregarding $2 \pi$) $ \frac{g^2 L}{p_{cut}}$ .

So the continuum limit does not always yield delocalization. For case 1 with
the regulator limit performed first (in the above sense) we obtain
localization. So even qualitatively the physics obtained is different.

\medskip
Let me finally mention that in light cone quantization (or quantization almost
on the light cone as in \cite{lenz}) performing first the limit to the light
cone corresponds to a weak coupling limit. Hence (even for case 2) none of the
dependence of the effective Hamiltonian on $\tilde c$ will be dominated by the
kinetic energy.

\section{Summary and Outlook }

I have shown that with case-1-sharp-cutoffs it is possible to obtain a
qualitatively different version of the Schwinger model in the continuum limit.
Namely, the zero mode can be dynamically frozen as opposed to the standard
delocalized version.

In $3+1$ dimensional QCD the dynamics of the zero modes is regarded to be
important for understanding low energy phenomena. One of the difficulties of
$3+1$ dimensional QCD is an extremely complicated dynamics of these zero modes
that are now $2+1$ dimensional fields rather than a single quantum mechanical
($0+1$ dimensional) variable. If a localization of the zero modes can be
obtained in a certain regularization, an interesting model could be formulated.
 As the dynamics of the zero modes is essentially altered, it might not have
much to do with QCD, but it may be easier to approach a solution of this model.

\medskip

I would like to thank F.Lenz for the supervision of this work and K. Yazaki for
the discussion about ordering of limits. I am grateful to the German Academic
Exchange  Service (DAAD) and the Bundesministerium f\"ur Forschung und
Technologie for supporting this work.


\begin{thebibliography}{99}
\bibitem{foot} I can assume that $c$ is not an integer, as the integers have
measure zero.
\end{thebibliography}
\end{document}